\documentclass{pasj00}
\draft
\usepackage{lineno} 
\usepackage{colortbl}
\usepackage{threeparttable}
\begin{document}
\SetRunningHead{S. Urakawa et al.}{Visible Spectroscopic Observations of Near-Earth Object 2012 DA$_{14}$}

\title{Visible Spectroscopic Observations of Near-Earth Object 2012 DA$_{14}$}

\author{Seitaro \textsc{Urakawa}} 
\affil{Bisei Spaceguard Center, Japan Spaceguard Association, 1716-3 Okura, Bisei, Ibara, Okayama 714-1411, Japan}
\email{urakawa@spaceguard.or.jp}
\author{Mitsugu \textsc{Fujii}}
\affil{4500 Kurosaki Tamashima Kurashiki Okayama 713-8126, Japan}\email{aikow@po.harenet.ne.jp}
\author{Hidekazu {\sc Hanayama}}
\affil{Ishigakijima Astronomical Observatory, National Astronomical Observatory of Japan, 1024-1 Arakawa, Ishigaki, Okinawa, 907-0024, Japan}\email{hanayama.hidekazu@nao.ac.jp}
\author{Jun {\sc Takahashi}}
\affil{Nishi-Harima Astronomical Observatory, Center for Astronomy, University of Hyogo, Sayo-cho, 679-5313, Hyogo, Japan}\email{takahashi@nhao.jp}
\author{Tsuyoshi {\sc Terai}}
\affil{National Astronomical Observatory, 2-21-1 Osawa, Mitaka, Tokyo 181-8588, Japan}\email{tsuyoshi.terai@nao.ac.jp}
\and
\author{Osamu {\sc Ohshima}}
\affil{Mizushima Technical High School, 1230 Nishiachi-cho, Kurashiki, Okayama 710-0807, Japan}\email{ohshima@po.harenet.ne.jp}


%


\maketitle

\begin{abstract}
We present visible spectroscopic observations of a near-earth object (NEO) 2012 DA$_{14}$. The asteroid 2012 DA$_{14}$ came close to the surface of the Earth on February 15, 2013 at a distance of 27,700~km. Its estimated diameter is around 45~m. The physical properties of such a small asteroid have not yet been well determined. The close encounter is a good opportunity to conduct a variety of observations. The purpose of this paper is to deduce the taxonomy of 2012 DA$_{14}$ by visible spectroscopic observations using the 0.4~m f/10 telescope at the Fujii Kurosaki Observatory.  We conclude that the taxonomy of 2012 DA$_{14}$ is an L-type in the visible wavelength region. In addition, we refer to the availability of a small, accessible telescope for NEOs smaller than 100~m.

\end{abstract}

\section{Introduction}
The asteroid taxonomy has been determined by a range of multi-band photometry and spectroscopy such as the Eight Color Asteroid Survey (ECAS) and Small Main-Belt Asteroid Spectroscopic Survey (SMASS) (e.g., \cite{Zel85}; \cite{Xu95}; \cite{Bus02a}; \cite{Bus02b}; \cite{Lazzaro04}; \cite{Leon10}).  Since the brightness increases when a near-earth object (NEO) is close to the Earth, the observations of NEO are suitable for the elucidation of taxonomy of sub-km-sized asteroids. \citet{Binzel04} analyzed the correlation for the populations, orbits, and diameter by the NEO taxonomy obtained. One result shows that Q-type NEOs are dominant in the range between 100~m and 5~km in diameter, compared with the prominence of S-type asteroids in ranges larger than 5~km. They suggested the reason was that the larger asteroids should have finer regolith size owing to their greater gravity and longer surface evolution lifetime, and alternatively the weaker effect of ``space weathering" for small asteroids due to the shorter collisional lifetimes. In addition to this, Earth encounters have been also thought to be the origin of fresh surfaces on NEOs \citep{Binzel10}. However, there are few spectroscopic observations for asteroids smaller than 100 m. \citet{Binzel04} has reported the taxonomy for a few asteroids smaller than 100~m. The taxonomy of 2012 KP$_{24}$ (diameter: 20~$\pm$~6~m) and 2012 KT$_{42}$ (diameter: 6~$\pm$~1~m) are C-type and B-type asteroids, respectively \citep{Pol12}. Most asteroids smaller than 100 m are monolithic asteroids. The constructions are different from the rubble-pile of sub-km-sized asteroids. To solve the collisional processes and mineral compositions for small asteroids, it is important to investigate whether the taxonomy has  correlation  between the construction and the diameter by examining gradual but steady observational data accumulations. 
 This study's purpose is to obtain the taxonomy for 2012 DA$_{14}$ at the time when the object moves fast on the sky after its closest approach to the Earth. 2012 DA$_{14}$ came close to the surface of the Earth on February 15, 2013 at a distance of 27,700~km. Its estimated diameter is around 45~m. The spectroscopic observations for such a small asteroid require a large-aperture telescope. \citet{Leon13} has reported that the taxonomy of 2012 DA$_{14}$ is an L-type by the visible spectra with visible and near-infrared color photometry using large-aperture telescopes. On the other hand, the close distance after its closest approach to the Earth assists us in conducting the spectroscopic observations using a small-aperture telescope if we have a skillful observational technique corresponding to the fast sky motion. We succeeded in the slit spectroscopic observations of 2012 DA$_{14}$ using the 0.4 m f/10 telescope. Our result contributes to the discussion for the rotational color variation.

\section{Observations}
We conducted the spectroscopic observation of 2012 DA$_{14}$ at the Fujii Kurosaki Observatory (FKO, Longitude = 133.6478$^{\circ}$ E, Latitude = 34.5100$^{\circ}$ N) with the Meade 0.4~m/f10 telescope on February 15, 2013. We used a FBSPEC-III spectroscope of our own construction. The FBSPEC-III is equipped with FLI ML6303E CCD with  3072 $\times$ 2048 pixels. The slit width and the provided spectral resolution are 5$^{\prime\prime}$ and $R$$\sim$500, respectively. This configuration permits the spectral range from 0.354~{$\mu$m} to 0.965~{$\mu$m}. Though we set an exposure time of six minutes, the fast sky motion of 2012 DA$_{14}$ made it difficult to hold the image within the slit width for the six-minute duration. When the sky motion is 1400$^{\prime\prime}$/min,  2012 DA$_{14}$ passes the slit width in 0.21~sec. Moreover, the transit time of 2012 DA$_{14}$ on the slit viewer is around eight seconds. Thus, we continued manual reintroduction of 2012 DA$_{14}$ into the slit every eight seconds for six minutes. By this procedure, we achieved the effective exposure time of around nine seconds for one frame. Since we obtained four frames, the total effective exposure time is around 36 seconds. Data reduction was conducted using the IRAF\footnote{IRAF is distributed by the National Optical Astronomy Observatory, which is operated by the Association of Universities for Research in Astronomy (AURA) under cooperative agreement with the National Science Foundation.} , under the following procedures. We carried out the subtraction by a dark image and division by averaged flat image. We had stocked more than 200 spectrum data for stars at the FKO. Wavelength calibrations were performed using a neon lamp. In addition, we also used the telluric absorption lines and the Balmer lines that were obtained from the past observation of a spectrophotometric standard star HR 1544 on September 22, 2011. The data of HR 1544 can be used because FBSPEC-III does not cause a wavelength gap with maximum of 0.3 nm in a different observational day experientially. The gap is negligible for the determination of asteroid taxonomy. Flux calibrations were conducted by using a spectrophotometric standard star HR 4963 observed at roughly the same airmass on the same nights. To obtain the relative reflectance of 2012 DA$_{14}$ with the enough accuracy, the spectrum of 2012 DA$_{14}$ was divided by the averaged spectrum of five solar analog stars, $\rho$ CrB,  $\mu^{2}$ Cnc, $\iota$ Per, $\lambda$ Aur, and  $\psi^{5}$ Aur. The spectra of five solar analog stars were also obtained by the past observations at the FKO, and were calibrated for the wavelength and flux. We obtained the averaged value of every 35 data, which corresponds to the wavelength interval of  $\sim$0.01~{$\mu$m}, as a representative reflectance of 2012 DA$_{14}$. However, we excluded  the data of around 0.73~{$\mu$m}, 0.76~{$\mu$m}, 0.82~{$\mu$m} and 0.91~{$\mu$m} where telluric H$_{2}$O and O$_{2}$ severely influence the spectrum of 2012 DA$_{14}$. Moreover, we eliminated the data in a wavelength region of shorter than 0.40~{$\mu$m} and longer than 0.92~{$\mu$m} because of the significant photometric error. Finally, we normalized the reflectance by the brightness at 0.55~{$\mu$m}. The observational states are listed in Table 1.

\begin{figure}
  \begin{center}
    \FigureFile(80mm,80mm){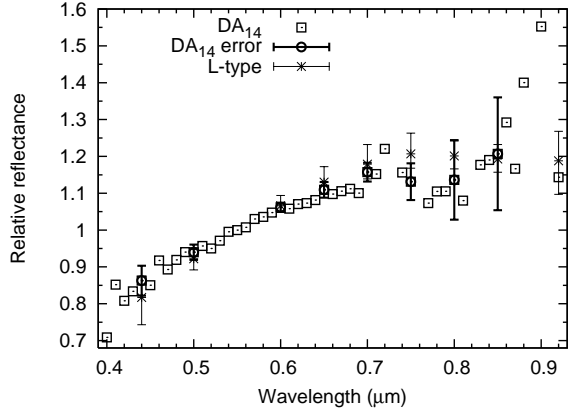}\\
  \end{center}
  \caption{Relative reflectance for 2012 DA$_{14}$.}\label{fig:1}
\end{figure}

\section{Result and discussion}
The relative reflectance for 2012 DA$_{14}$ is shown in Figure 1. Though a typical error for relative reflectance of 2012 DA$_{14}$ is around 0.02, the data with a lower integration time have the errors of over 0.10 in a wavelength region of shorter than 0.41~{$\mu$m} and longer than 0.78~{$\mu$m}. The legend of ``DA$_{14}$ error" in Figure 1 indicates the error at the representative wavelength. The error at 0.85~{$\mu$m} is calculated by stacking the data from 0.83~{$\mu$m} to 0.87~{$\mu$m}. The error at the 0.92~{$\mu$m} is omitted because the enough accuracy was not achieved even if the data are stacked. The taxonomy is determined by following the flow chart of \citet{Bus99}. Collaterally, we also compare the obtained relative reflectance with the typical relative reflectance for each taxonomy. As shown in Figure 1, there is no strong absorption in the wavelength region of longer than 0.80~{$\mu$m}.  Thus, we eliminate the taxonomy that exhibits the concave-up curvature in the wavelength regions. Then, we can narrow the search to K-, L-, or Ld-type when the relative reflectance at 0.75~{$\mu$m} is more than 1.10 and less than 1.35. We obtain the relative reflectance of 1.13 $\pm$ 0.05 at 0.75~{$\mu$m}. Next, the overall slope of spectrum from 0.44~$\mu$m to 0.92~$\mu$m is calculated by the least-square fit to the equation:
\begin{equation}
r = 1.0+\gamma(\lambda-0.55),
\end{equation}
where $r$ is the relative reflectance at every 0.01~{$\mu$m}, $\lambda$ is the wavelength in microns, and $\gamma$ is the overall slope of the fitted line normalized to have the value of unity at 0.55~{$\mu$m}. When we assume the relative reflectance has same error at each wavelength to ensure effective use of data that is obtained in the wavelength region of longer than 0.80~{$\mu$m}, the overall slope is $\gamma$ = 0.86 $\pm$ 0.07. The taxonomy is an L-type when the $\gamma$ is more than 0.6 and less than 0.9. The taxonomy is the Ld-type when the $\gamma$ is more than 0.9 and less than 1.35. Moreover, we defined the deviation between the observational value and the representative relative reflectance for each taxonomy as follows,   
  \begin{equation}
\sigma = \sqrt{\left(\frac{\sum^{9}_{i = 1}\left(O_{i}-E_{i}\right)^{2}}{9}\right)},
\end{equation}
where $i$ = 1, 2, $\cdots$, 8, 9 corresponds to the relative reflectance at 0.44~{$\mu$m}, 0.50~{$\mu$m}, 0.60~{$\mu$m}, 0.65~$\mu$m, 0.70~{$\mu$m}, 0.75~{$\mu$m}, 0.80~{$\mu$m}, 0.85~$\mu$m, and 0.92~{$\mu$m}. $O_{i}$ and $E_{i}$ are the $i$-th observational relative reflectance and the $i$-th relative reflectance of the Bus taxonomy, respectively. The results are summarized in Table 2. We conclude that taxonomy of 2012 DA$_{14}$ is an L-type because the relative reflectance at 0.75~{$\mu$m} and the overall slope satisfy the requirements for L-type. Moreover, the deviation of Ld-type is significantly larger than that of L-type. The legend of ``L-type" in Figure 1 shows the range of relative reflectance for L-type asteroids. The observational results are included within the range of L-type. However, we have to take notice the possibility of T-type when the relative reflectance at 0.75~$\mu$m is lower than 1.10. Our results are consistent with the result of \citet{Leon13} in the visible wavelength regions. The consistency indicates that 2012 DA$_{14}$ has not shown the apparent rotational color variation because our observational time is different from that of \citet{Leon13}. We also conducted the photometric observations, the polarization observations, and the near-infrared observations for 2012 DA$_{14}$ at the Saitama University telescope and Nishi-Harima Astronomical Observatory. The rotational period (Terai et al., in preparation), the polarization degree and the near-infrared photometry (Takahashi et al., in preparation) make clearer the physical properties of 2012 DA$_{14}$. 

The discovery of close-approaching NEOs smaller than 100 m is increasing due to the progress of spaceguard activities, such as Pan-STARRS \citep{Ka02}, Catalina Sky Survey \citep{Lar98}, Spacewatch \citep{Car94} , NEOWISE \citep{Mai11}, etc. In the case of 2012 DA$_{14}$, a number of observations have been carried out all over the world because of  the accurate ephemeris prediction. For the discussion of the rotational color variation of 2012 DA$_{14}$, multi-band photometry and/or spectroscopic observations are required at various locations. This study contributes to the resolution of surface properties at the time when 2012 DA$_{14}$ moves fast on the sky after its closest approach to the Earth. On the other hand, the physical properties of NEOs smaller than 100 m have not been well determined because the opportunities for observation are limited immediately after their discovery. Immediate observation systems by small and accessible telescopes such as this study help us to elucidate the physical properties of NEOs smaller than 100 m.


\begin{table*}
\begin{center}
\caption{Observation states.\label{tbl-1}}
\begin{threeparttable}
\begin{tabular}{ccccccc}
\hline
Object &  Year/Mon/Day (UT) & EXP time(s) &  Airmass & $\alpha$ [$^{\circ}$]\tnote{a}& Sky motion [$''$/min]\\
\hline
2012 DA$_{14}$ & 2013/ 2/15.86435 & $\sim$9 & 1.444 &35.88& 1561\\
2012 DA$_{14}$ & 2013/ 2/15.86854 & $\sim$9 & 1.458 &37.96 & 1435 \\
2012 DA$_{14}$ & 2013/ 2/15.87272 & $\sim$9 & 1.474 &39.90 & 1321\\
2012 DA$_{14}$ & 2013/ 2/15.87685 & $\sim$9 & 1.491 &41.70 & 1218\\
\hline
\end{tabular}
\begin{tablenotes}
\item[a]Phase angle (Sun--2012 DA$_{14}$--observer).
\end{tablenotes}
\end{threeparttable}
\end{center}
\end{table*}

\begin{table}
\begin{center}
\caption{Standard deviation $\sigma$ for each taxonomy. The taxonomy is arranged in ascending order for the $\sigma$. \label{tbl-2}}
\begin{threeparttable}
\begin{tabular}{cc}
\hline
Taxonomy &  $\sigma$\\

\hline
T,  K, Xk, L & $\textless$  0.05 \\ \hline
Xe, S, Sl, D, Sk, X, &0.05 -- 0.10 \\
Sa, Xc, Sr, Cg, Sq, Ld & \\ \hline
C, A, R, Cb, Cgh, & $\textgreater$  0.10 \\ 
Ch, Q, B, V, O&    \\ \hline

\end{tabular}
\end{threeparttable}
\end{center}
\end{table}

\bigskip


\end{document}